\documentclass[twocolumn]{aastex631}

\usepackage{booktabs}  
\usepackage{longtable} 
\usepackage{graphicx} 
\usepackage{wrapfig}
\usepackage{placeins}  
\usepackage{soul}         
\sethlcolor{yellow}
\usepackage{hyperref}
\usepackage{lineno}
\usepackage{xcolor}

\begin{document}

\title{A Comprehensive Catalog of Radio Sources and Rotation Measures in the Perseus Molecular Cloud from Very Large Array Observations}


\author{Haleh Hajizadeh}
\affiliation{Department of Physics and Astronomy, The University of Calgary, 2500 University Drive NW, Calgary AB T2N 1N4, Canada}
\email{haleh.hajizadeh@ucalgary.ca}

\author{Jeroen Stil}
\affiliation{Department of Physics and Astronomy, The University of Calgary, 2500 University Drive NW, Calgary AB T2N 1N4, Canada}

\author{René Plume}
\affiliation{Department of Physics and Astronomy, The University of Calgary, 2500 University Drive NW, Calgary AB T2N 1N4, Canada}

\author{Mehrnoosh Tahani}
\affiliation{Department of Physics and Astronomy, The University of South Carolina, Columbia, 29208, SC, USA}

\author{Preshanth Jagannathan}
\affiliation{National Radio Astronomy Observatory, Socorro, NM 87801, USA}

\begin{abstract}
We present a comprehensive radio polarization study of the Perseus molecular cloud using wideband L-band observations from the Karl G. Jansky Very Large Array. Our survey covers $\sim13.8$~deg$^2$ with a mean Stokes~$I$ sensitivity of $\sim80~\mu$Jy~beam$^{-1}$, enabling the detection of 1410 compact radio sources. From this population, we construct a catalog of source properties, including positions, integrated flux densities, and spectral indices measured across nine spectral windows. The majority of sources exhibit negative spectral indices, consistent with non-thermal synchrotron emission.
Using RM Synthesis and RM CLEAN techniques, we detect 205 polarized background sources above an $8\sigma$ threshold. This corresponds to a sampling density of $\sim14.8$~deg$^{-2}$, representing more than an order-of-magnitude increase compared to previous NVSS-based measurements. The resulting rotation measures exhibit coherent large-scale variations across the surveyed region, with additional small-scale structure superimposed.
The enhanced sensitivity, frequency coverage, and sampling density of our observations enable a substantially improved mapping of the line-of-sight magnetic field component toward the Perseus molecular cloud compared to previous surveys.
\end{abstract}



\section{Introduction} \label{sec:intro}

Stars form within molecular clouds that are often organized into elongated filaments and dense clumps. Studies have shown that magnetic fields influence the internal structure and evolution of these regions by controlling how filaments fragment into dense cores, providing additional pressure support against gravitational collapse, and guiding the large-scale morphology and orientation of the gas \citep{crutcher2012magnetic, mckee2007theory, hennebelle2019role, krumholz2019role, Pattle2023, tahani2022three, liu2022magnetic}. Despite significant progress, the relative importance of magnetic fields compared with turbulence and gravity remains under debate. In some scenarios, magnetic pressure can stabilize clouds and slow their collapse, whereas in others turbulence or gravity dominates when the field is weak or highly disordered \citep{hennebelle2019role, mignon2023role, suin2025role}. Addressing how magnetic fields interact with turbulence and gravity to shape the formation and evolution of stars, therefore, requires sensitive, multi-wavelength observations capable of tracing magnetic structure across spatial scales \citep{tahani2022orion, tahani20223d}.

Magnetic fields in molecular clouds can be investigated using a variety of observational techniques, each sensitive to different components of the field. While the plane-of-sky component has been extensively mapped and analyzed (e.g., \cite{tahani2023jcmt, hwang2025jcmt}), constraints on the line-of-sight component ($B_\parallel$) remain comparatively limited. Probing $B_\parallel$ requires different diagnostics.
One method for probing magnetic fields in molecular clouds is through Zeeman splitting, which provides measurements of the magnetic field strength in dense regions; however, such observations are technically challenging and extremely sparse \citep{crutcher2019review}.
Faraday rotation of polarized radio emission, in contrast, is significantly more efficient for sampling the magnetized medium and provides valuable information about the three-dimensional structure of the field along the line-of-sight \citep{brentjens2005faraday, burn1966depolarization}. 
In this case, the observed polarization angle (electric vector position angle; EVPA) \(\Psi(\lambda)\) varies with wavelength as  $\Psi(\lambda) = \Psi_0 + \Phi\,\lambda^2$,
 where \(\Psi_0\) is the intrinsic polarization angle, \(\lambda\) is the wavelength [m],  and \(\Phi\) [rad\,m\(^{-2}\)] is the \textit{Faraday depth}, which quantifies the cumulative effect of Faraday rotation along the line-of-sight. It encodes information about $B_\parallel$ and the distribution of free electrons through the relation

\begin{equation}
\label{equation:faraday}
\Phi(l) = 0.812 \int_{Source}^{Observer} n_e(l')\, B_{\parallel}(l')\, dl' ,
\end{equation}

where \(B_{\parallel}\) [\(\mu\)G] is the line-of-sight magnetic field strength, \(n_e\) [cm\(^{-3}\)] is the free electron density, and \(dl\) [pc] is the differential path length through the ionized medium. The quantity \(\Phi(l)\) represents the \textit{Faraday depth} accumulated up to position \(l\), while the observed \textit{Rotation Measure (RM)} corresponds to the net Faraday depth integrated over the full line-of-sight to the polarized source, and is equal to \(\Phi\) only in the case of a single foreground Faraday-rotating screen.

Using the Faraday rotation, \cite{tahani2018helical} developed a technique called MC-BLOS to map $B_\parallel$ across molecular clouds. For the Perseus molecular cloud, they found a $B_\parallel$ sign reversal across the filamentary cloud and used Planck dust polarization data to infer the plane-of-sky magnetic field component. From the observer's point of view, their data showed a concave, arc-shaped magnetic field, offering insights into how magnetic fields influence molecular clouds and star formation \citep{tahani2018helical, tahani2019could, tahani2024mc}.

MC-BLOS makes use of the all-sky RM catalog derived from NVSS observations \citep{taylor2009rotation}. While this catalog provided the first large-scale map of Faraday rotation across the sky, it is subject to several well-known limitations. The relatively low sensitivity of NVSS results in a sparse sampling of polarized background sources, with a surface density of approximately one polarized source per deg$^{-2}$. Moreover, RMs in this catalog were derived using only two frequency channels, leading to $n\pi$ ambiguities and large uncertainties in the inferred $B_\parallel$ \citep{van2011modeling}. 
To mitigate these issues, MC-BLOS applied strict uncertainty thresholds that further reduced the usable sample size. These limitations underscore the need for the higher-sensitivity, broadband RM surveys we present here.

\begin{figure*}[]
\centering
\includegraphics[width=0.8\textwidth]{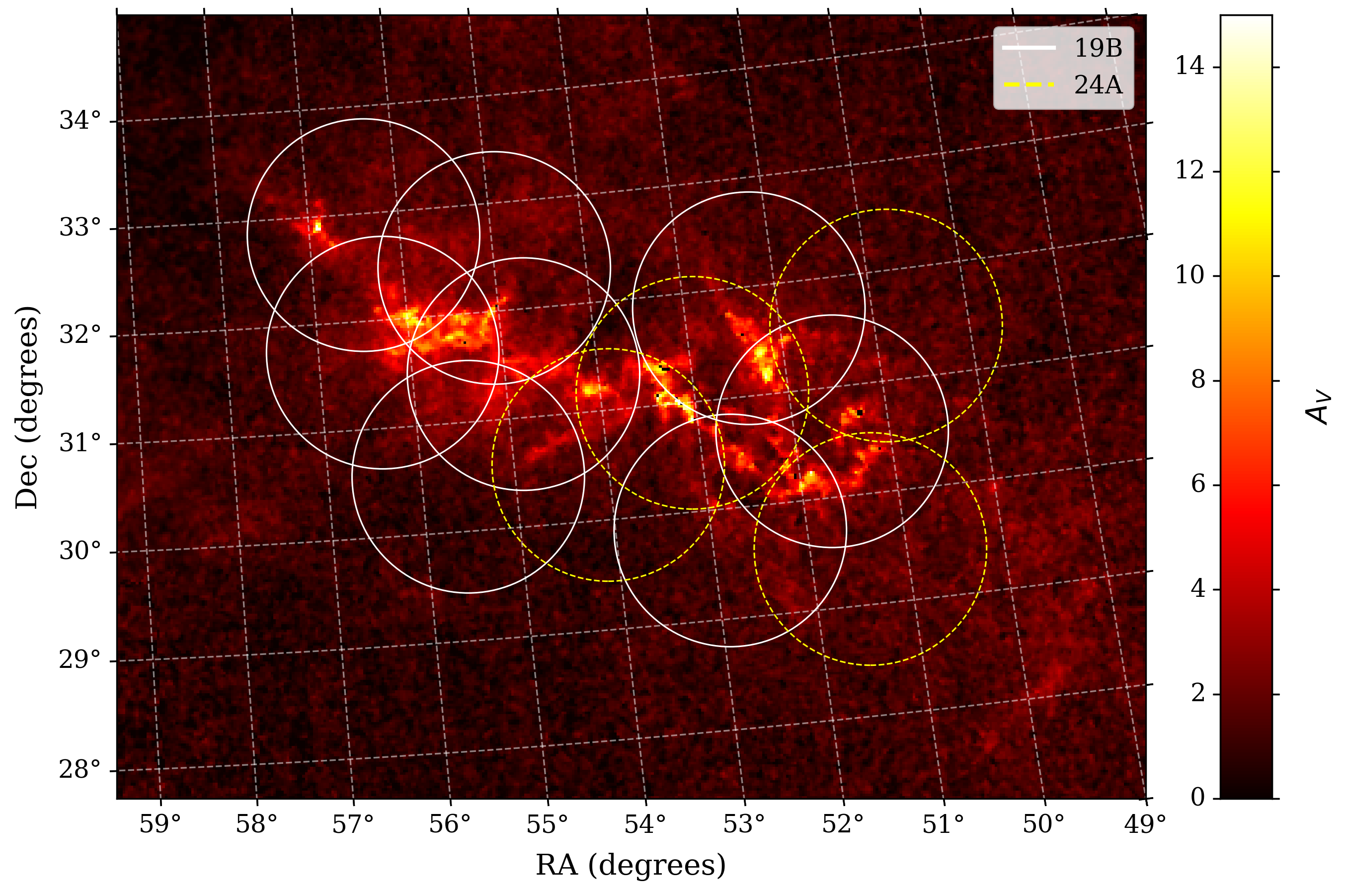}
\caption{A$_V$ map of the Perseus molecular cloud. Each circle represents a mosaic composed of 55 fields arranged in a hexagonal pattern, with each mosaic observed over five epochs. The four numbers of yellow dashed circles represent the mosaics observed with the VLA in 2019, forming the primary dataset used in this study. The eight white circles indicate the additional regions observed in 2024, aimed at enhancing the completeness of the $B_\parallel$ measurements across the cloud. The color scale in this figure represents the visual extinction ($A_V$, in magnitudes) derived from near-infrared dust extinction maps based on data from the Two Micron All Sky Survey (2MASS) and processed using the NICEST algorithm \citep{kainulainen2009probing, lombardi2009nicest}.}
\label{fig:perseuscloud}
\end{figure*}

In this paper, we present a broadband L-band polarization survey of the Perseus cloud conducted with the Karl G. Jansky Very Large Array (VLA).\footnote{The National Radio Astronomy Observatory is a facility of the National Science Foundation operated under cooperative agreement by Associated Universities, Inc.}

The Perseus molecular cloud is an ideal candidate for studying magnetic fields due to its relative proximity of $294 \pm 17$~pc \citep{vallenari2023gaia}. Located in the constellation Perseus, it serves as a key laboratory for probing cloud evolution and early stellar development. Moreover, the Perseus cloud exhibits an intriguing magnetic field morphology, with $B_\parallel$ reversing direction across the cloud. It has been mapped extensively in dust continuum and molecular lines \citep[e.g.,][]{ridge2006complete, enoch2006bolocam, curtis2010submillimetre}, revealing complex structures such as filamentary networks, molecular outflows, and shells.
Recent Green Bank Telescope ammonia observations have traced filament accretion and fragmentation down to $\sim 0.05$~pc scales \citep{chen2024filament}, while chemical surveys have revealed a rich inventory of complex organic molecules in starless and pre-stellar cores \citep{scibelli2024survey}.
In the radio regime, VLA C-band (4–8~GHz) interferometric surveys \citep{tobin2016vla, tychoniec2017vla} have provided detailed census data of compact protostellar sources but were not optimized for tracing diffuse emission or polarized background sources. High-resolution, broadband RM measurements are therefore essential for tracking $B_\parallel$ gradients and relating magnetic structure to filaments and feedback features \citep{tahani20223d}.

We compile a comprehensive, high-sensitivity ($\sim80~\mu$Jy~beam$^{-1}$) catalog of background extragalactic radio sources, reporting integrated and peak flux densities across nine spectral windows (SPW) and deriving their spectral indices. This represents the most complete catalog of radio sources toward the Perseus molecular cloud to date. From this population, we construct a refined catalog of RMs, characterized via broadband Faraday synthesis, for polarized background sources. The survey's broad $\lambda^2$ coverage, combined with RM Synthesis \citep{brentjens2005faraday, rudnick2024pseudo,han2017observing, purcell2020rm} and RM CLEAN \citep{heald2017rm} deconvolution, mitigates the $n\pi$ ambiguities inherent to two-frequency surveys like NVSS and suppresses sidelobe contamination and bandwidth depolarization. The resulting RM catalog is both higher in source density and more reliable than previous compilations \citep[e.g.,][]{taylor2009rotation}, providing a robust foundation for mapping $B_\parallel$ across the Perseus cloud. The enhanced sensitivity reveals numerous compact, polarized background sources, resulting in a tenfold improvement in RM sampling density across the cloud.

\begin{figure*}
\centering 
\includegraphics[width=0.8\textwidth, height=0.6\textwidth]{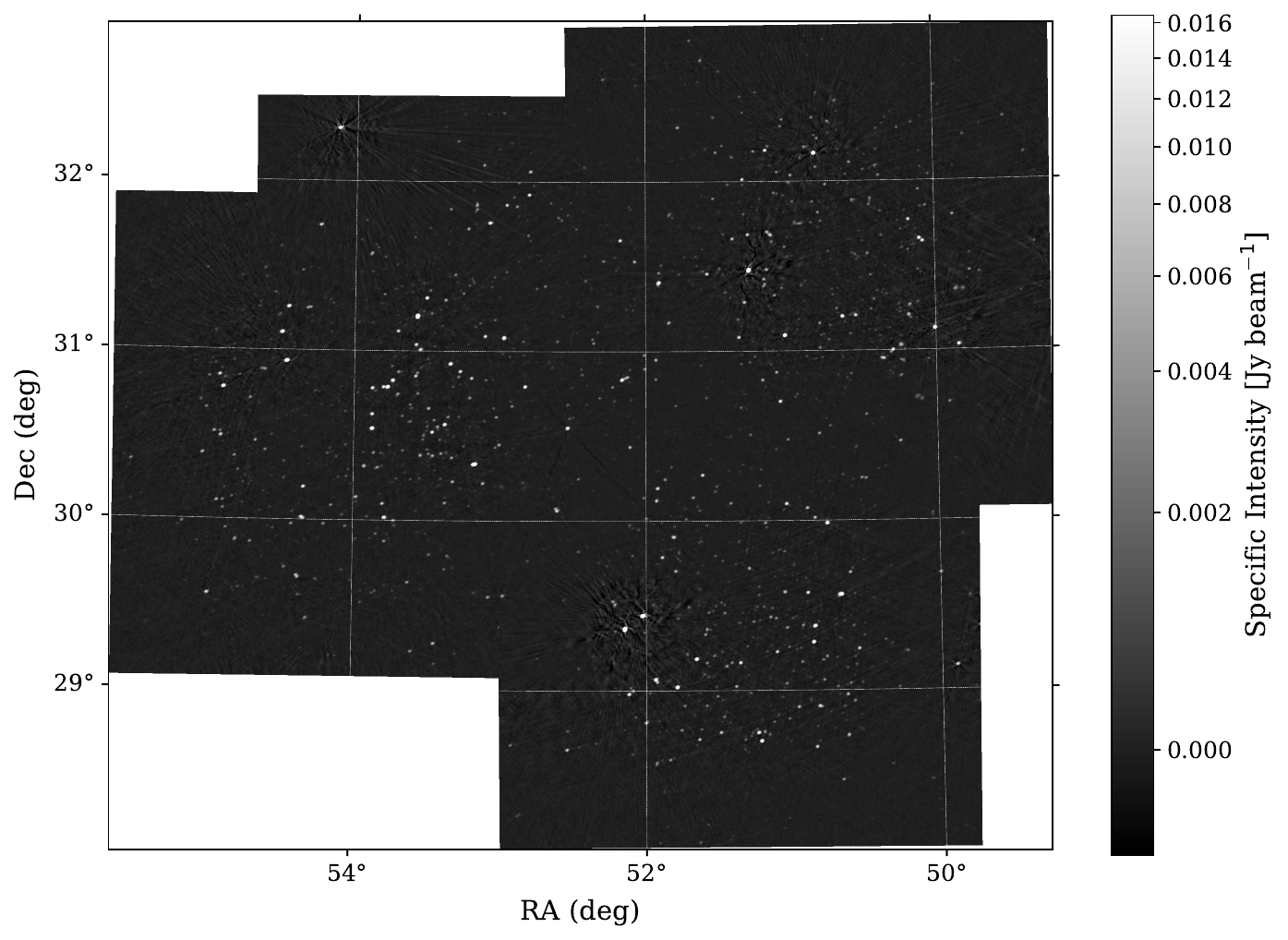}
\caption{Stokes I image of the Perseus molecular cloud obtained from the 19B VLA observations in D-configuration. This mosaic represents the total intensity map produced using the multi-frequency synthesis imaging technique with wide-field corrections and deconvolution applied via the \texttt{tclean} algorithm in CASA. The image covers the central region of the molecular cloud with a synthesized beam size of approximately 46$^{\prime\prime}$ and a pixel scale of 2.5$^{\prime\prime}$. The enhanced sensitivity achieved in this dataset ($\sim80~\mu$Jy beam$^{-1}$) allows for the detection of numerous faint background radio sources, which are critical for probing the magnetic field structure via Faraday rotation.}
\label{fig:bigmosaic} 
\end{figure*}

The structure of this paper is as follows: Section~\ref{sec:data} describes the VLA observations and data reduction steps, including calibration, imaging, and polarization analysis. Source extraction, spectral index fitting, and catalog validation are presented in Sections~\ref{results}. We conclude in Section~\ref{sec:conclusion} with a summary and outlook for future work.

\section{Observations and Data Reduction} \label{sec:data}

\subsection{Observation} \label{observation}

\begin{figure*}[ht]

\centering
\includegraphics[width=0.8\textwidth, height=0.4\textwidth]{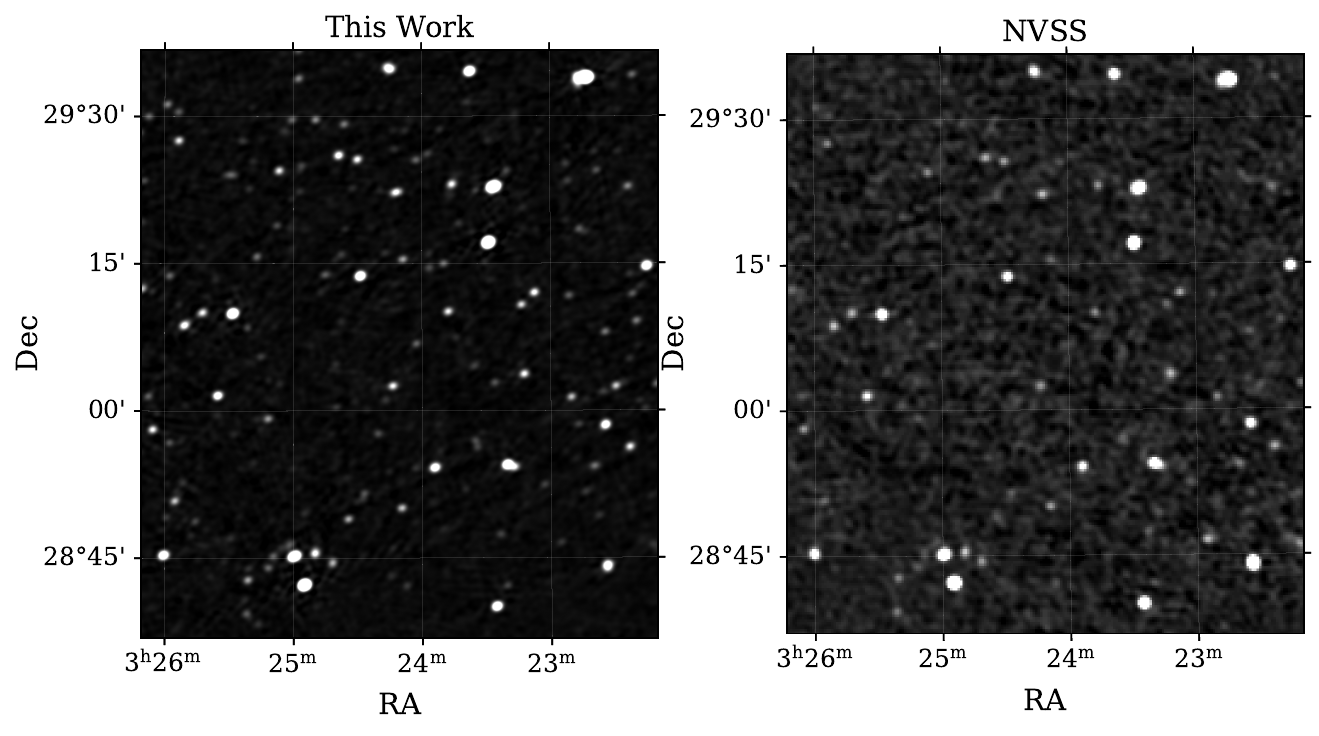}
\caption{
Comparison between a $1^{\circ} \times 1^{\circ}$ region from this work (left) and the corresponding area in NVSS (right), shown on the same intensity scale. Owing to the higher sensitivity and angular resolution of our observations, the left panel reveals a substantially larger population of compact radio sources. In this region, we detect 120 sources using a $5\sigma$ source-extraction threshold, compared to 34 sources identified in NVSS.
}

\label{fig:comparison}

\end{figure*}

The observations for this study were conducted using the VLA. Figure~\ref{fig:perseuscloud} shows the extinction map of the Perseus cloud \citep{kainulainen2009probing}, illustrating how our observed regions trace the structure of the cloud. The mosaic fields were arranged in a hexagonal grid pattern consisting of 55 pointings designed to provide uniform sensitivity across the target region. Each field overlaps slightly with its neighbors to ensure continuous sky coverage over an approximately circular area with a diameter of about 2.2 deg$^2$, with an integration time of $\sim$2~minutes per field.

There were two separate VLA observing programs covering the Perseus molecular cloud: program 19B-053, obtained in 2019, and program 24A-376, approved as follow-up observations in 2024 (hereafter referred to as the 19B and 24A datasets, respectively). Together, these observations span the region roughly \(49^\circ < \mathrm{RA} < 56^\circ\) and \(28^\circ < \mathrm{Dec} < 33^\circ\), as shown in Figure~\ref{fig:perseuscloud}. The fields observed in 2019, marked by yellow circles in Figure \ref{fig:perseuscloud}, constitute the primary dataset analyzed in this study. The additional fields, observed in 2024 and indicated by white circles, were designed to improve spatial coverage and increase the density of RM measurements across the Perseus cloud. In this paper, we focus on the 19B dataset; the analysis of the 24A follow-up observations will be presented in a forthcoming publication.

The 19B dataset was obtained using the VLA D-configuration, the most compact configuration.
The synthesized beam is elliptical and varies across the band, ranging from
$60\arcsec \times 43\arcsec$ at the lowest frequency to
$34\arcsec \times 30\arcsec$ at the highest frequency (FWHM major $\times$ minor axes).
The observations covered four mosaic regions, each observed over five separate epochs to improve the signal-to-noise ratio (SNR) and sampling of the visibility plane.

The spectrometer was set up to cover the whole L-band (930-2070 MHz) in full polarization mode. The data were split into 18 SPWs, each with a bandwidth of 64 MHz. There were 64 frequency channels in each SPW with a channel nominal width of 1 MHz. 
The use of L-band is especially advantageous for RM research because it provides an optimal balance between sensitivity to Faraday rotation and depolarization effects, allowing polarized emission to be detected while still preserving information on the intervening magneto-ionic medium \citep{burn1966depolarization, brentjens2005faraday}.

\subsection{Data Reduction} \label{sec:data-reduc}

The data were reduced using the Common Astronomy Software Applications (CASA) package \citep{mcmullin2007casa, casa2022casa} following a standard interferometric workflow that included calibration, imaging, and polarization analysis. The calibration procedure consisted of flux density and bandpass calibration, as well as polarization calibration to correct for instrumental leakage and to set the absolute polarization angle. A significant fraction of the visibilities was affected by radio frequency interference (RFI); this was mitigated using CASA’s \texttt{flagdata} task with both \texttt{tfcrop} and \texttt{rflag} modes, followed by manual inspection to refine the flagged data. The \texttt{tfcrop} mode identifies outliers in time--frequency space, while \texttt{rflag} flags statistical outliers based on deviations from local time and frequency averages. These steps were essential for preserving astrophysical signal integrity and ensuring reliable downstream analysis.

Following the flagging process, 9 of the 18 original SPWs were retained for subsequent imaging. Spectral windows were considered usable when residual RFI was confined to narrow sub-bands and the post-flagging noise was comparable to adjacent SPWs. SPWs were excluded when contamination remained broadband across a large fraction of the band and/or when the post-flagging noise was substantially elevated. These clean SPWs are centered at $1:1345.5883$, $2:1409.5883$, $3:1473.5883$, $4:1653.5753$, $5:1717.5753$, $6:1781.57534$, $7:1845.5753$, $8:1909.5753$, and $9:2037.5753~$ MHz, each with a bandwidth of $64~$ MHz.

The VLA calibration pipeline \citep{nrao_vla_pipeline} was used to do standard calibration, which included calibrating the flux, bandpass, and gain. The primary flux calibrator was 3C147, adopting the Perley–Butler 2017 model \citep{perley2013accurate}, which provides a frequency-dependent flux density scale widely used for VLA observations. Bandpass and polarization angle calibration were performed using 3C138 \citep{perley2013integrated}, while J0336+3218 served as the gain and phase calibrator. Polarization calibration, which is not handled by the standard pipeline, was performed manually. We derived leakage terms using 3C147 as an unpolarized calibrator, which has been shown to have stable total flux density and negligible polarization across the L-band in recent years.


\subsection{Imaging and Deconvolution} \label{Imaging}

After calibration, the visibility data were imaged using the CASA \texttt{tclean} algorithm, which reconstructs the sky brightness distribution from the interferometric measurements. The multi-scale multi-frequency synthesis (MT-MFS) deconvolver \citep{rau2011multi} was employed to fit for spectral index variations within the field.

We used the following imaging parameters to get the best image quality and sensitivity: 

\begin{itemize} 

\item Gridder: \texttt{awproject}, which takes into consideration wide-field effects and baselines that are not coplanar \citep{bhatnagar2013wide}. 

\item Weighting Scheme: Briggs weighting with \texttt{robust} = 0.5 \citep{briggs1995high}, balancing sensitivity and angular resolution. 

\item Cell Size and Image Dimensions: The images were reconstructed with a pixel scale of 2.5$''$ and an image size of 4096 pixels per side to fully sample the sub-mosaic. This pixel scale was chosen to provide fine sampling for source detection and to improve the positional accuracy of detected sources, while the synthesized beam sets the effective angular resolution.

\item Threshold and Stopping Criteria: The cleaning threshold was set at 100 $\mu$Jy, ensuring that faint sources were properly deconvolved while avoiding over-cleaning artifacts. 

\end{itemize}

The final cleaned Stokes $I$ mosaic for the 19B dataset is shown in Figure \ref{fig:bigmosaic}. The image covers the central region of the Perseus cloud with a synthesized beam of 46$''$, a primary-beam FWHM of 32.14$'$, and a pixel scale of 2.5$''$. In addition to revealing numerous faint background radio sources, crucial for Faraday rotation studies, the mosaic also exhibits several imaging artifacts associated with bright sources. These include radial sidelobe patterns and spokes that trace the VLA's configuration uv-coverage, as well as localized regions with elevated residual structure around the brightest continuum emitters. Such features are expected in wide-field L-band mosaics when bright sources lie near the edge of individual pointings. Although these patterns do not affect the identification of compact background sources, they are noted here for completeness and should be considered when interpreting local variations in the noise properties across the mosaic. 

Our observations achieve an average sensitivity of 80~$\mu$Jy~beam$^{-1}$, defined here as the background RMS noise measured in source-free regions of the final Stokes~$I$ images. In the vicinity of the brightest continuum sources (within $\sim 5'$), the local RMS increases from the field-average value to typically $\sim$\,140~$\mu$Jy~beam$^{-1}$ (up to $\sim$\,200~$\mu$Jy~beam$^{-1}$ in the most affected regions).
This sensitivity is more than a factor of five better than the 450~$\mu$Jy~beam$^{-1}$ sensitivity of the NVSS survey of the Perseus molecular cloud \citep{condon1998nrao}. This substantial improvement in sensitivity directly enhances our ability to detect faint sources that were previously undetected in NVSS.
As shown in Figure \ref{fig:comparison}, our dataset identified approximately 120 sources within a region of $1^{\circ} \times 1^{\circ}$ centered at $(51.04583^{\circ}, 29.11333^{\circ})$. In the same field, the NVSS survey detected only 34 sources. Thus, our observations revealed about 3.5 times more sources than NVSS, confirming how higher sensitivity enables the detection of fainter radio sources.

\subsection{Calculating Rotation Measure} \label{sec:rm_synthesis}

As discussed in Section~\ref{sec:intro}, Faraday rotation provides a powerful probe of $B_\parallel$ in ionized astrophysical environments. 
In the idealized case of a single Faraday-thin screen, the polarization angle varies linearly with $\lambda^2$, and the observed RM corresponds to a single Faraday depth component. In realistic astrophysical environments, however, polarized emission often probes multiple Faraday-rotating regions along the line-of-sight, including the emitting source, the interstellar medium (ISM), and the Galactic foreground. The resulting polarization signal is therefore a superposition of multiple Faraday components, for which a simple linear fit of polarization angle versus $\lambda^2$ is inadequate.

The combined effect of internal and external Faraday screens leads to wavelength-dependent depolarization through several mechanisms. When polarized synchrotron emission originates throughout a volume and experiences different amounts of rotation along the line-of-sight, the superposition of these depth-dependent polarization states produces depth depolarization. This effect is distinct from beam depolarization, which arises from unresolved spatial fluctuations in Faraday depth within the synthesized beam, and from bandwidth depolarization, which occurs when polarization angles vary significantly across a finite frequency channel \citep{burn1966depolarization, sokoloff1998depolarization, van2017faraday}.

To overcome these limitations, we employ RM Synthesis, a Fourier-based technique that reconstructs the distribution of polarized emission as a function of Faraday depth along the line-of-sight \citep{brentjens2005faraday}. RM Synthesis exploits the Fourier relationship between the complex polarization,
$P(\lambda^2)=Q(\lambda^2)+i\,U(\lambda^2)$, and the Faraday dispersion function (FDF), $F(\phi)$. In this work, the FDF was recovered using discrete, weighted RM Synthesis over the sampled $\lambda^2$ values, explicitly accounting for flagged channels and gaps in frequency coverage. The resulting FDF represents the convolution of the intrinsic Faraday spectrum with the rotation measure spread function (RMSF). To mitigate the effects of this convolution, we apply RM CLEAN to deconvolve the RMSF and recover the underlying Faraday structure, enabling robust identification of dominant Faraday depth components and characterization of Faraday-complex behavior along the line-of-sight.

\subsection{Cube Construction for RM Synthesis}
\label{cube}

To extract RMs from our data, we followed a multi-step approach:

\begin{itemize} 

\item Each 64~MHz-wide SPW was first averaged over eight consecutive 1~MHz channels, resulting in eight 8~MHz-wide bins per SPW. For each 8~MHz-wide bin, we imaged the Stokes~$I$, $Q$, and~$U$ parameters using the \texttt{tclean} task in CASA. The imaging used \texttt{gridder=mosaic} to account for multiple pointings in the mosaic, and we adopted a consistent image size of 4096~pixels with a cell size of 2.5\arcsec.

\item Imaging was performed with Briggs weighting (robust = 0.5) and a maximum of 5000 iterations per image, using a cleaning threshold of 100 $\mu$Jy. We applied primary beam correction (\texttt{pbcor=True}) and used \texttt{mosweight=True} to account for sensitivity variations across the mosaic.

\item The imaging process was repeated for each of the nine SPWs, resulting in one Stokes~$Q$ and one Stokes~$U$ image per 8~MHz segment. In total, $9$~SPWs~$\times$~$8$~frequency bins produced $72$ images for the final cube.

\item To ensure all images within the cube have a uniform angular resolution, \texttt{imsmooth} was applied to convolve each image to a common restoring beam corresponding to the largest synthesized beam among the spectral channels (60$^{\prime\prime}$). Smoothing to the lowest-resolution beam ensures that all frequency slices sample the same spatial scales, preventing artificial frequency-dependent variations in the Stokes $Q$ and $U$ spectra that could create false polarization signals.

\item These images were then concatenated to construct data cubes for each Stokes parameter, allowing the extraction of polarization spectra at each spatial pixel. 

\item The resulting Stokes Q and U spectra were analyzed using RM Synthesis to reconstruct the FDF. The analysis was carried out using the RM-Tools package, which implements both RM Synthesis and RM CLEAN.

\end{itemize}

In the following, we indicate the three fundamental parameters of the RM Synthesis framework following the formalism described in \citet{brentjens2005faraday}. 

The resolution in Faraday space, determined by the full width at half maximum (FWHM) of the RMSF, is approximately $\delta\phi \approx 98~\mathrm{rad~m^{-2}}$ for our dataset. This implies that Faraday components separated by less than this value cannot be reliably distinguished. 

The maximum observable Faraday depth, beyond which strong depolarization occurs within a single channel, depends on the minimum channel width in wavelength-squared space. For our data, which has a channel width corresponding to 8 MHz, this yields a maximum detectable Faraday depth of about $|\phi_{\text{max}}| \approx 3.3 \times 10^{3}~\mathrm{rad~m^{-2}}$ at a frequency of 1.4 GHz. This ensures sensitivity to large RMs (typically expected only within the disk and toward the Galactic center) without significant bandwidth depolarization \citep{takahashi2021introduction}. 

Finally, the maximum Faraday thickness, which defines the largest scale in Faraday depth to which the data remain sensitive, is set by the shortest sampled wavelength-squared value. For our observations, this corresponds to approximately $\phi_{\text{max-scale}} \approx 145~\mathrm{rad~m^{-2}}$, establishing the upper limit on the width of Faraday-thick structures detectable before they are depolarized by differential Faraday rotation. 

Together, these parameters---resolution, maximum observable depth, and maximum scale---define the range of Faraday depths and scales in magneto-ionic structure that can be probed using RM Synthesis.

\section{Results \& Discussion}
\label{results}
\subsection{Radio Source Extraction} \label{sec:pybdsf}

\begin{figure}
\centering
\hspace*{-1.0cm}
\includegraphics[width=0.55\textwidth, height=0.4\textwidth]{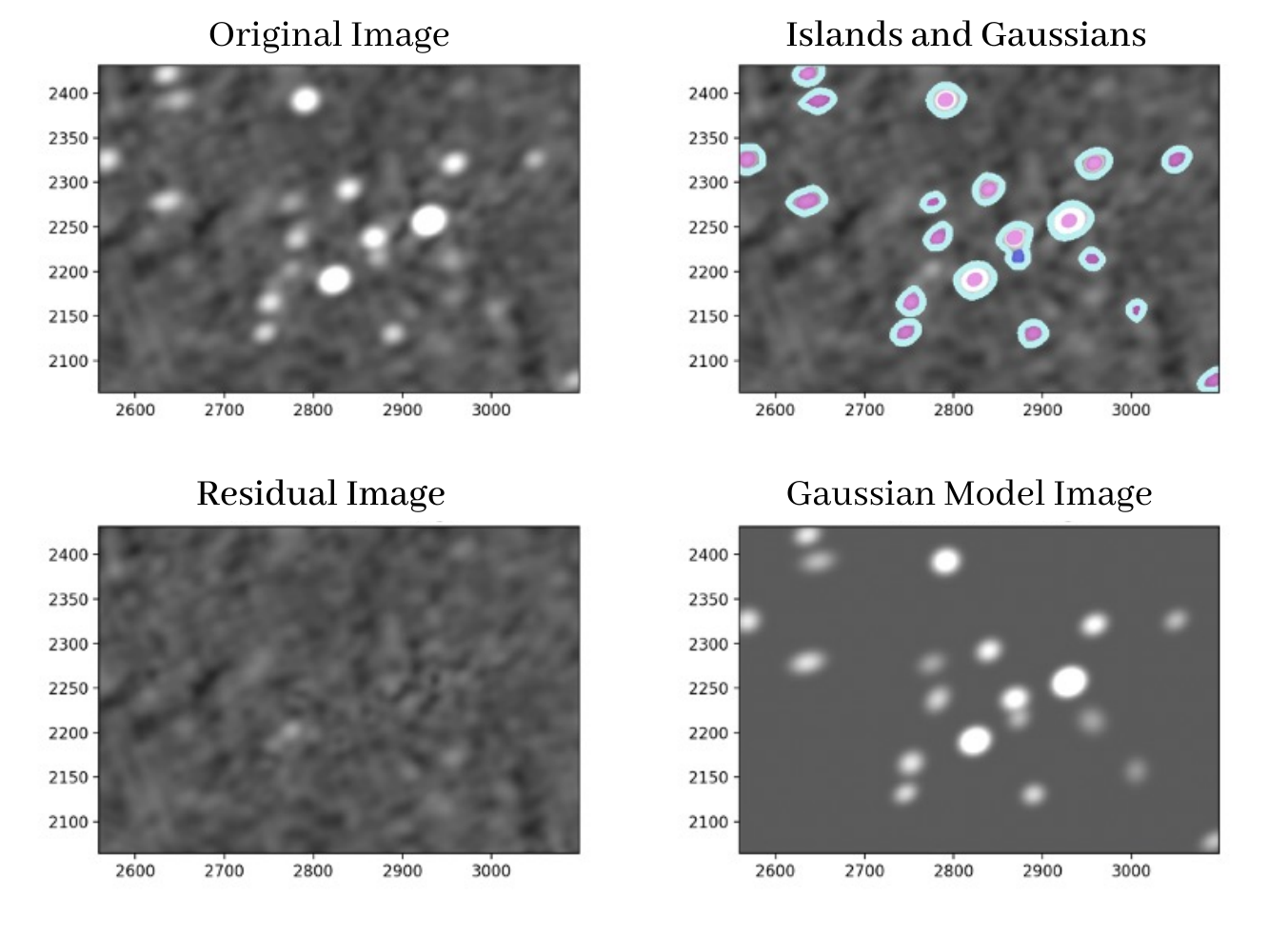}
\caption{Example output from \texttt{PyBDSF} showing source detection and Gaussian modeling. 
\textbf{Top-left:} Original Stokes $I$ image displayed with a logarithmic stretch to emphasize faint emission. 
\textbf{Top-right:} Detected emission islands outlined in cyan. Individual Gaussian components fitted within each island are shown as magenta ellipses. Blue ellipses represent the final fitted sources, which may consist of one or more Gaussians. 
\textbf{Bottom-left:} Residual image obtained by subtracting the model from the original image, ideally containing only noise. The remaining bright pixels correspond to emission below the $5\sigma$ threshold.
\textbf{Bottom-right:} Gaussian model image synthesized from all fitted components, with white regions indicating modeled emission. 
}
\label{fig:pybdsf_example}
\end{figure}

We used PyBDSF (Python Blob Detector and Source Finder), a widely used algorithm designed for the extraction and characterization of radio sources in interferometric data \citep{mohan2015pybdsf}. The algorithm is particularly suitable for processing large datasets.


\tabletypesize{\scriptsize}
\begin{deluxetable*}{cccc}

\tablecaption{This table provides the label, units, and description for each column in the full multi-SPW source catalog. It serves as a descriptive metadata summary for the complete dataset, which includes integrated and peak flux densities, spectral indices, associated uncertainties, and local RMS noise across all nine SPWs. Due to the large number of columns, the full table is not shown in print but is available as a MRT file in the electronic edition of the {\it Astrophysical Journal}.}

\label{tab:sourcecatalog}
\tablehead{
  \colhead{column number} & \colhead{Units} & \colhead{Label} & \colhead{Description}
}
\startdata
1 & \nodata & Source\_ID & Unique source ID \\
2 & deg & RA & Right ascension (J2000) \\
3 & deg & DEC & Declination (J2000) \\
4 & deg & E\_RA & Error in RA \\
5 & deg & E\_DEC & Error in DEC \\
6 & \nodata & alpha & Spectral index \\
7 & \nodata & E\_alpha & Error in spectral index \\
8 & mJy & Total\_S\_SPW1 & Integrated flux density in SPW1 \\
9 & mJy & E\_Total\_S\_SPW1 & Error in integrated flux density in SPW1 \\
10 & mJy & Peak\_S\_SPW1 & Peak flux density in SPW1 \\
11 & mJy & E\_Peak\_S\_SPW1 & Error in peak flux density in SPW1 \\
12 & mJy & RMS\_SPW1 & Island RMS noise in SPW1 \\
13 & mJy & Total\_S\_SPW2 & Integrated flux density in SPW2 \\
14 & mJy & E\_Total\_S\_SPW2 & Error in integrated flux density in SPW2 \\
15 & mJy & Peak\_S\_SPW2 & Peak flux density in SPW2 \\
16 & mJy & E\_Peak\_S\_SPW2 & Error in peak flux density in SPW2 \\
17 & mJy & RMS\_SPW2 & Island RMS noise in SPW2 \\
18 & mJy & Total\_S\_SPW3 & Integrated flux density in SPW3 \\
19 & mJy & E\_Total\_S\_SPW3 & Error in integrated flux density in SPW3 \\
20 & mJy & Peak\_S\_SPW3 & Peak flux density in SPW3 \\
21 & mJy & E\_Peak\_S\_SPW3 & Error in peak flux density in SPW3 \\
22 & mJy & RMS\_SPW3 & Island RMS noise in SPW3 \\
23 & mJy & Total\_S\_SPW4 & Integrated flux density in SPW4 \\
24 & mJy & E\_Total\_S\_SPW4 & Error in integrated flux density in SPW4 \\
25 & mJy & Peak\_S\_SPW4 & Peak flux density in SPW4 \\
26 & mJy & E\_Peak\_S\_SPW4 & Error in peak flux density in SPW4 \\
27 & mJy & RMS\_SPW4 & Island RMS noise in SPW4 \\
28 & mJy & Total\_S\_SPW5 & Integrated flux density in SPW5 \\
29 & mJy & E\_Total\_S\_SPW5 & Error in integrated flux density in SPW5 \\
30 & mJy & Peak\_S\_SPW5 & Peak flux density in SPW5 \\
31 & mJy & E\_Peak\_S\_SPW5 & Error in peak flux density in SPW5 \\
32 & mJy & RMS\_SPW5 & Island RMS noise in SPW5 \\
33 & mJy & Total\_S\_SPW6 & Integrated flux density in SPW6 \\
34 & mJy & E\_Total\_S\_SPW6 & Error in integrated flux density in SPW6 \\
35 & mJy & Peak\_S\_SPW6 & Peak flux density in SPW6 \\
36 & mJy & E\_Peak\_S\_SPW6 & Error in peak flux density in SPW6 \\
37 & mJy & RMS\_SPW6 & Island RMS noise in SPW6 \\
38 & mJy & Total\_S\_SPW7 & Integrated flux density in SPW7 \\
39 & mJy & E\_Total\_S\_SPW7 & Error in integrated flux density in SPW7 \\
40 & mJy & Peak\_S\_SPW7 & Peak flux density in SPW7 \\
41 & mJy & E\_Peak\_S\_SPW7 & Error in peak flux density in SPW7 \\
42 & mJy & RMS\_SPW7 & Island RMS noise in SPW7 \\
43 & mJy & Total\_S\_SPW8 & Integrated flux density in SPW8 \\
44 & mJy & E\_Total\_S\_SPW8 & Error in integrated flux density in SPW8 \\
45 & mJy & Peak\_S\_SPW8 & Peak flux density in SPW8 \\
46 & mJy & E\_Peak\_S\_SPW8 & Error in peak flux density in SPW8 \\
47 & mJy & RMS\_SPW8 & Island RMS noise in SPW8 \\
48 & mJy & Total\_S\_SPW9 & Integrated flux density in SPW9 \\
49 & mJy & E\_Total\_S\_SPW9 & Error in integrated flux density in SPW9 \\
50 & mJy & Peak\_S\_SPW9 & Peak flux density in SPW9 \\
51 & mJy & E\_Peak\_S\_SPW9 & Error in peak flux density in SPW9 \\
52 & mJy & RMS\_SPW9 & Island RMS noise in SPW9 \\
\enddata

\end{deluxetable*}

PyBDSF starts by creating a noise map to capture how the background noise varies across the image, which helps separate real sources from noise. It then identifies groups of pixels that stand out, potential sources, and fits them with Gaussian models to measure their position, size, shape, and brightness. Figure~\ref{fig:pybdsf_example} illustrates the PyBDSF output for a portion of our field, showing the detected sources along with their fitted Gaussian models.

The final step in the PyBDSF process is catalog creation. A comprehensive catalog of the detected sources, including their fitted parameters and associated uncertainties, is generated. To construct it, we used Stokes $I$ images integrated over each of the nine SPWs separately, and the catalog reports the flux values measured in these SPWs. This catalog serves as the foundation for further analysis (see Tables \ref{tab:sourcecatalog} and \ref{tab:spw15-matched}, which are described in Section \ref{radio_catalog}). 

\subsection{Catalog of Radio Sources} \label{radio_catalog}

To facilitate a comprehensive understanding of our source catalog, we provide two summary tables. Table~\ref{tab:sourcecatalog} offers a detailed description of the complete machine-readable table (MRT) that comes with this publication. It lists the label, units, and definition for each column in the catalog, which includes source positions, spectral indices, integrated and peak flux densities, associated uncertainties, and local RMS noise measurements across all nine SPWs. Given the large number of parameters, 52 columns in total, the full table is made available in MRT format in the electronic edition of the Astrophysical Journal, as it is too extensive to be included in print.


\subsubsection{Position and Flux Comparison with NVSS}

To evaluate the positional and flux reliability of our catalog, we compared our source detections with the NVSS catalog \citep{condon1998nrao}. We focused on the SPW2 map (centered at 1409.5883~MHz with a bandwidth of 64~MHz), since it closely matches the NVSS observing frequency of 1.4~GHz. 

We carried out the comparison within a circular region of radius 0.8$^\circ$ centered on the mosaic’s phase center; this area was selected because it corresponds to the region of highest sensitivity in our observations. Within this region, we detected 480 sources in SPW2 and found 368 sources in the NVSS catalog by computing the angular separation between sources and using a matching radius of 30$''$. This radius is appropriate because both datasets have an angular resolution of $\sim$45$''$ \citep{condon1998nrao}, so small position differences between the two catalogs are expected. Using 30$''$, about two-thirds of the beam size, allows us to account for these typical shifts while still keeping the chance of false matches low. Of these, 148 sources in our SPW2 map had no NVSS counterpart, while 36 NVSS sources did not appear in our data. These differences can be explained by several factors: minor positional shifts caused by differences in observing frequency and resolution, as well as possible imaging artifacts or noise that may have prevented source detection in one dataset or the other. The NVSS sources not recovered in our data are predominantly low-flux-density detections, suggesting that sensitivity and survey differences are the primary drivers of the mismatch rather than intrinsic variability.

\begin{figure}[]
\centering
\includegraphics[width=0.49\textwidth, height=0.4\textwidth]{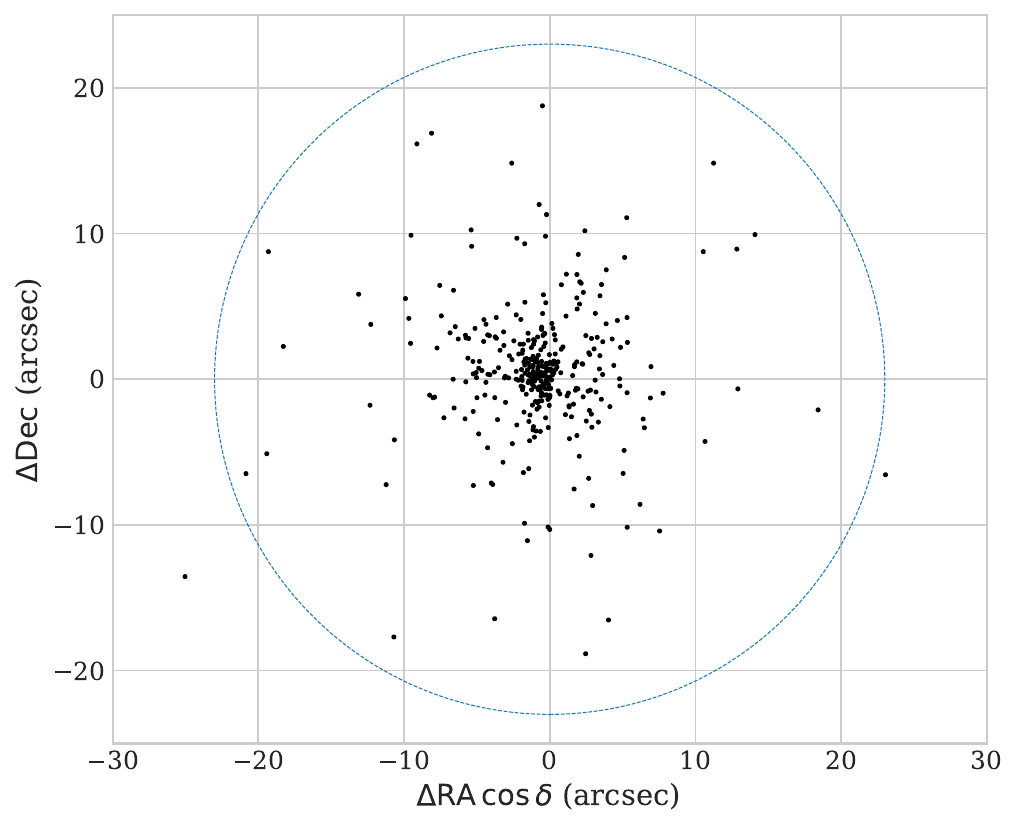}
\caption{Positional offsets in RA and DEC between matched sources in our VLA catalog and a reference catalog. 
Each point represents the angular offset $(\Delta \mathrm{RA}, \Delta \mathrm{DEC})$ for a matched source. 
The distribution is tightly clustered around the origin, indicating overall good astrometric agreement, with most offsets well within one synthesized beam. 
The mean offsets are $\langle \Delta \mathrm{RA} \rangle = -1.\!\!^{\prime\prime}160 \pm 0.\!\!^{\prime\prime}305$ and 
$\langle \Delta \mathrm{DEC} \rangle = 0.\!\!^{\prime\prime}512 \pm 0.\!\!^{\prime\prime}273$. 
A few sources show larger deviations (up to $\sim 30^{\prime\prime}$).}
\label{fig:positional_offset}
\end{figure}

To evaluate the reliability of our source positions, we compared the source positions to those in the NVSS. The offsets in RA and DEC are shown in Figure~\ref{fig:positional_offset}. The mean image offsets are $\langle \Delta \mathrm{RA} \rangle = -1.\!\!^{\prime\prime}160 \pm 0.\!\!^{\prime\prime}305$ and, $\langle \Delta \mathrm{DEC} \rangle = 0.\!\!^{\prime\prime}512 \pm 0.\!\!^{\prime\prime}273$ which is smaller than the size of a pixel (2.5$''$). The distribution is tightly clustered around the origin, with most offsets well within the synthesized beam (46$''$), indicating good overall agreement between the two catalogs. A small number of sources display larger deviations (up to $\sim$ 30$''$), which are likely associated with extended emission, source blending, or low SNR detections.

\begin{figure}[]
\centering
\includegraphics[width=0.46\textwidth, height=0.34\textwidth]{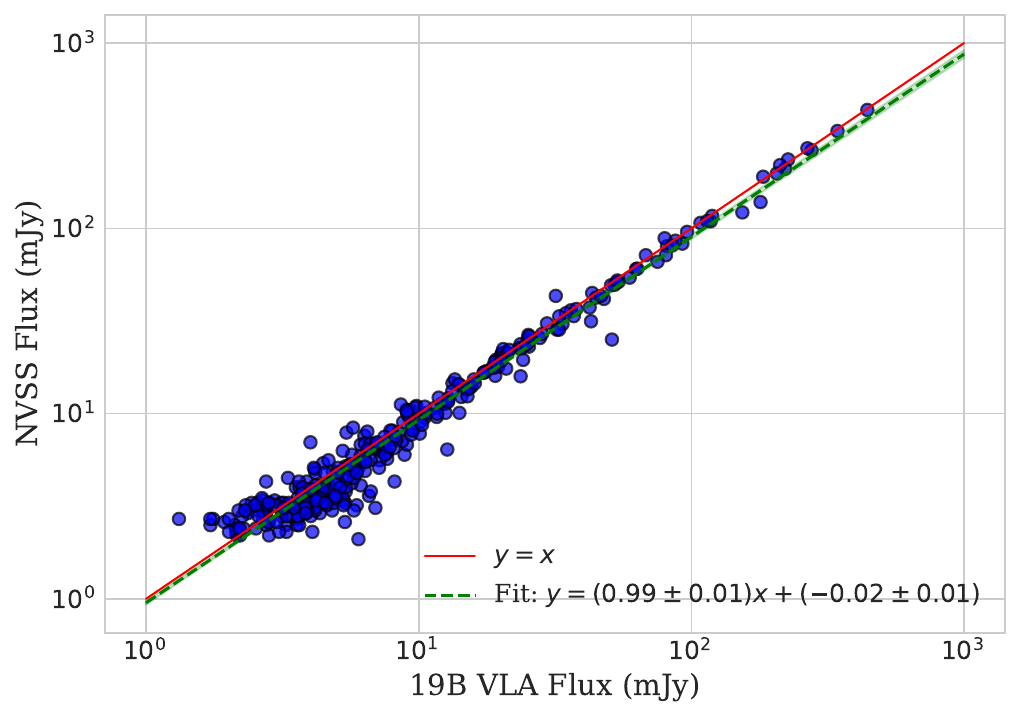}
\caption{Comparison of total flux densities from our dataset and NVSS for sources detected in both surveys. The red line represents the one-to-one correspondence. The strong linear relationship confirms the consistency between the two datasets.}
\label{fig:flux_comparison}
\end{figure}

To check how reliable our flux measurements are, we compared the total flux densities of sources found in both our catalog and the NVSS. The results are shown in Figure~\ref{fig:flux_comparison}, where we plot the total flux densities measured in our data against those reported in the NVSS. This comparison checks to see if our flux calibration is correct and shows that the two surveys generally agree with each other. The red line marks the one-to-one relation, while the green dashed line shows a linear fit to the data. The close alignment between the two indicates that our flux calibration is consistent with NVSS. Small deviations from the exact correspondence can be attributed to differences in sensitivity limits or intrinsic source variability over the decades separating the surveys. Overall, the strong linear correlation demonstrates that our flux scale is well tied to previous radio surveys, providing confidence that our dataset is robust for subsequent RM Synthesis and polarization analyses.

\begin{figure}
\centering
\includegraphics[width=0.48\textwidth, height=0.34\textwidth]{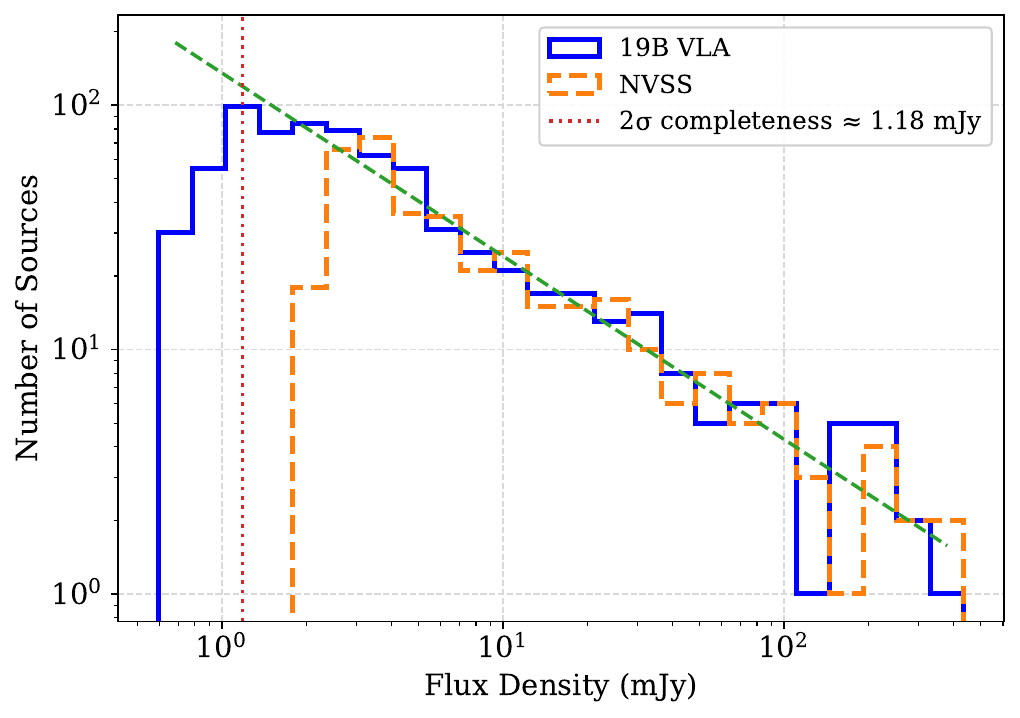}
\caption{
Log--log histogram of integrated flux densities for radio sources within the custom mask region toward the Perseus cloud. The solid blue line corresponds to our 19B VLA catalog, while the dashed orange line shows the NVSS source counts over the same area. The VLA observations recover substantially more faint sources, reflecting their lower noise level and improved sensitivity relative to NVSS. The green dashed curve represents the expected slope, and the vertical red dotted line marks the estimated 2$\sigma$ completeness limit of our catalog at $\approx 1.18$~mJy, determined by the flux density at which the observed counts begin to fall below the extrapolated power-law trend.}
\label{fig:log_flux_histogram}
\end{figure}

To further assess the flux distribution and sensitivity of our catalog, we constructed the differential source counts shown in Figure~\ref{fig:log_flux_histogram}. The VLA catalog (solid blue line) contains significantly more faint sources than the NVSS (dashed orange line), reflecting the much lower noise level of our 19B observations. Whereas NVSS becomes increasingly incomplete below $\sim 3$--$4$~mJy, our catalog continues to recover sources well into the sub-mJy regime. To quantify this, we estimated the completeness by identifying the flux level at which the observed counts fall systematically below the extrapolated slope (green dashed line). The turnover occurs at $\sim$1.18~mJy, which we adopt as the 2$\sigma$ completeness threshold (vertical dotted red line). This value is fully consistent with the nominal sensitivity of our maps and represents the flux density above which our catalog contains all detectable point sources across the surveyed region. The deeper completeness of our VLA observations compared to NVSS ensures that our sample includes a much larger population of faint background sources.

\subsection{Spectral Indices} \label{spectral_index}

After source extraction, we used the spectral index ($\alpha$) to characterize the dominant radio emission mechanisms. The spectral index is defined through the power-law relation,
\begin{equation}
S_{\nu} \propto \nu^{\alpha},
\end{equation}
where $S_{\nu}$ is the flux density at the observing frequency $\nu$.

For each source, the spectral index was determined by fitting a power-law model to the measured flux densities across the available SPWs using the \texttt{lmfit} non-linear least-squares minimization package \citep{newville2016lmfit}. The fit was performed with inverse-variance weighting, using the squared flux-density uncertainties.

The uncertainty on the spectral index was obtained from the covariance matrix returned by the fitting procedure. Specifically, the reported uncertainty corresponds to the square root of the diagonal covariance element associated with $\alpha$, assuming Gaussian measurement errors and local linearity of the model near the best-fit solution. This value represents the formal $1\sigma$ uncertainty on the spectral index.

Negative spectral indices (\(\alpha < -0.1\)) are characteristic of synchrotron radiation, the dominant emission mechanism in extragalactic radio sources and active galactic nuclei (AGN). In contrast, flat or positive spectral indices (\(\alpha \gtrsim 0\)) can arise from several physical processes. Partially optically thick thermal bremsstrahlung (free--free) emission from ionized gas produces moderately positive indices (e.g., \(\alpha \sim 0.6\)), as observed in stellar winds and compact H\,\textsc{ii} regions \citep{wright1975radio}. Rising spectra may also result from absorption effects, such as synchrotron self-absorption or free--free absorption, which suppress low-frequency emission; this behavior is typical of GHz-peaked spectrum and compact steep-spectrum sources that exhibit a low-frequency spectral turnover \citep{pacholczyk1971radio, o1998compact}.

\begin{figure}
\centering
\includegraphics[width=0.46\textwidth, height=0.34\textwidth]{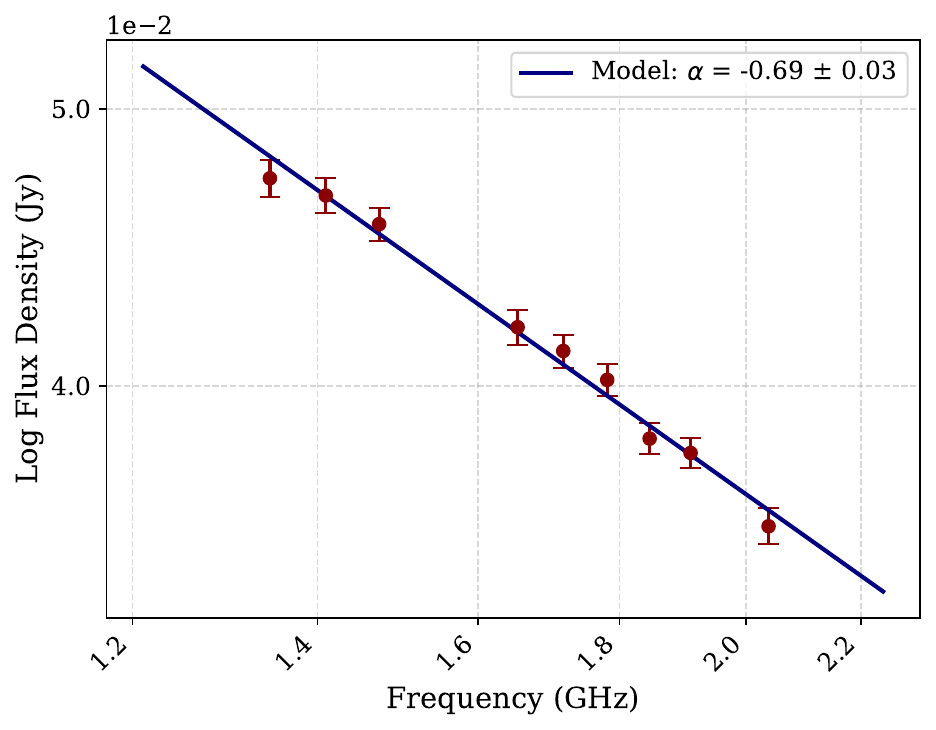}
\caption{Spectral fit for a representative radio source. The data points correspond to flux densities measured in each SPW, and the blue line shows the best-fit power-law model. The derived spectral index for this source is $\alpha = -0.69 \pm 0.03$, consistent with synchrotron emission.}
\label{fig:spectral_index}
\end{figure}

\begin{figure}
\centering
\includegraphics[width=0.48
\textwidth, height=0.36\textwidth]{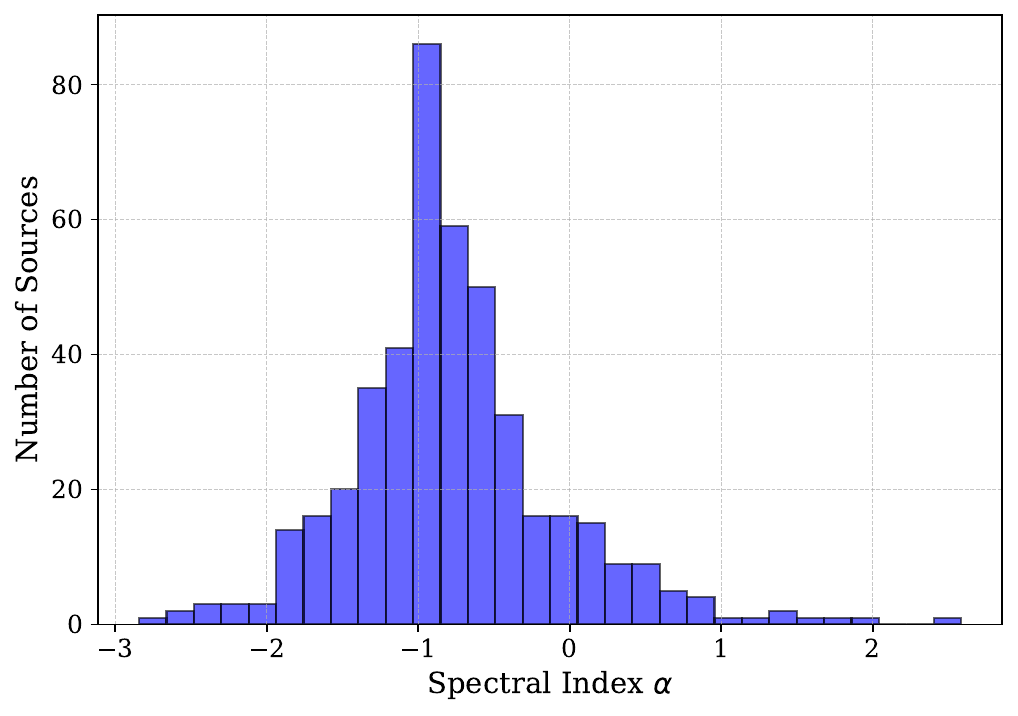}
\caption{
Histogram of spectral indices, $\alpha$, for radio sources in the catalog. The distribution has a mean spectral index of $\langle \alpha \rangle = -0.77$ and a median value of $\tilde{\alpha} = -0.85$.
}
\label{fig:spectral_indices}
\end{figure}

\begin{figure*}
\centering
\includegraphics[width=1\textwidth]{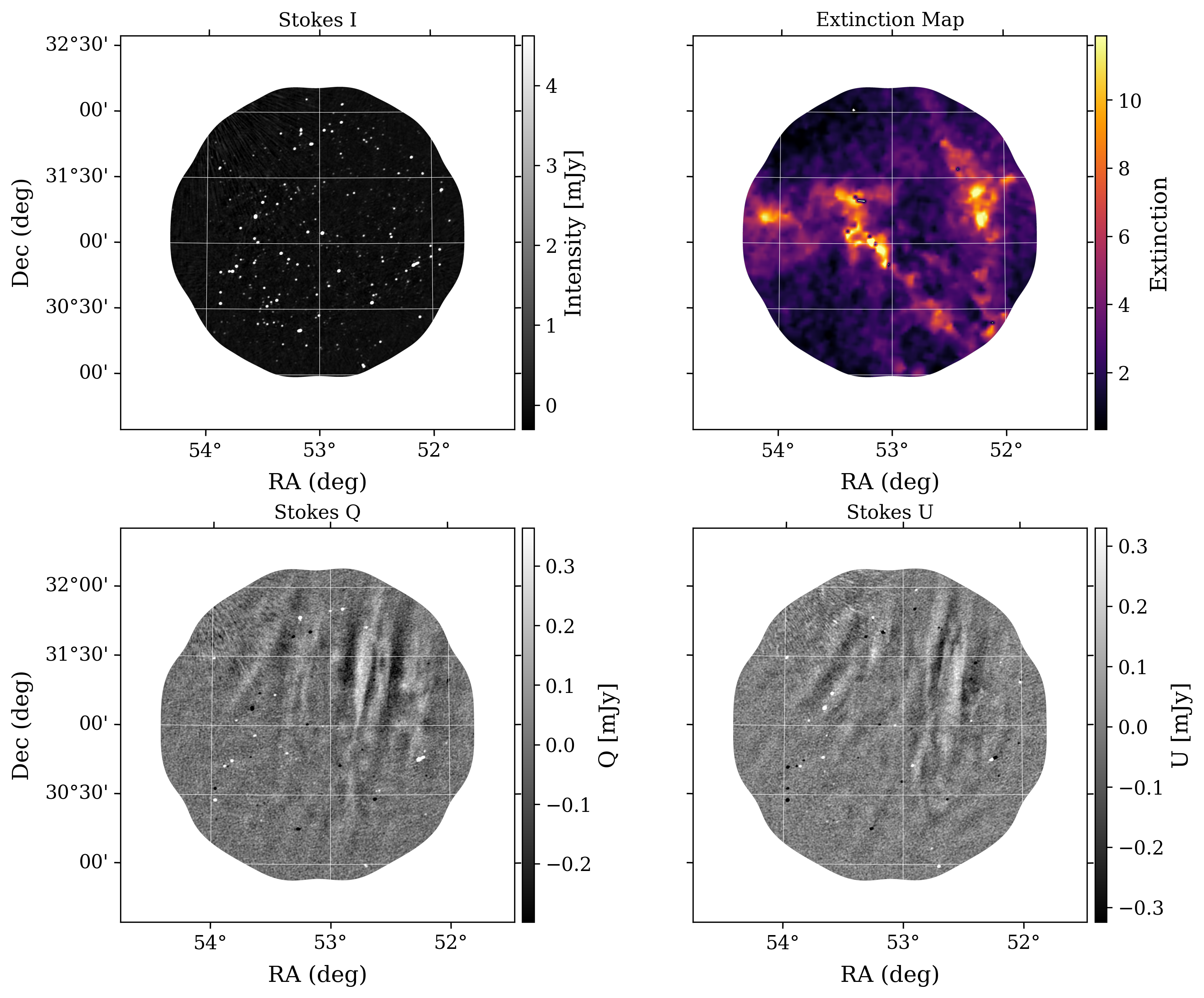}
\caption{
Stokes $I$ (top left), Stokes $Q$ (bottom left), and Stokes $U$ (bottom right) images for a representative mosaic centered on the Perseus molecular cloud, alongside an extinction map \citep{kainulainen2009probing} (top right) of the same region. The Stokes $I$ panel highlights compact and diffuse radio continuum sources, while the extinction map traces the underlying dust structure of the molecular cloud. Blank pixels in the extinction map correspond to regions where stellar measurements were excluded to minimize bias from foreground stars and stars embedded within the cloud, as well as from sources rejected through an iterative sigma-clipping procedure at the $5\sigma$ level relative to the local extinction field. The large-scale fluctuations in Stokes $Q$ and $U$ arise from diffuse Galactic synchrotron emission, whose polarization structure is further modified by Faraday rotation in the intervening magneto-ionic medium.
}
\label{fig:stokesQU}
\end{figure*}

Figure \ref{fig:spectral_index} illustrates an example of spectral index fitting for one of the detected radio sources in our catalog. The plot shows the measured total flux densities across multiple frequency bands and the best-fit power-law model overlaid. The spectral index $\alpha$ is derived from a log-log linear fit to the data. In this example, the source at RA = 53.19709$^\circ$, DEC = 30.83951$^\circ$ has a spectral index of $\alpha = -0.69 \pm 0.03$, indicating that synchrotron emission dominates the radio emission.

Figure~\ref{fig:spectral_indices} shows the distribution of spectral indices, $\alpha$, for radio sources in our catalog with reliable flux–frequency measurements. To ensure robust spectral index estimates, we restrict the analysis to sources detected in at least four SPWs and located within regions of high sensitivity in the mosaics, applying a geometric mask defined as a circular region of radius $0.8^\circ$ centered on the mosaic phase center. This criterion excludes edge sources where primary-beam correction becomes large and flux uncertainties increase. In addition, we require the reduced chi-square of the power-law fit to satisfy $\chi^2_{\nu} < 3$, selecting well-constrained spectral fits. The distribution is dominated by negative spectral indices, consistent with synchrotron emission from extragalactic radio sources. The mean spectral index is $\langle \alpha \rangle = -0.77$, while the median value is $\tilde{\alpha} = -0.85$, with a standard deviation of $0.68$.


\subsection{Galactic Foreground Emission in the Data} \label{galactic_foreground}

To provide context for the impact of polarized foreground emission on our analysis, Figure~\ref{fig:stokesQU} illustrates a sample Stokes $I$, $Q$, and $U$ mosaics together with the corresponding extinction map for a region in the Perseus field centered at RA = 53.01888$^\circ$ and DEC = 31.08444$^\circ$. The top-left panel shows the Stokes $I$ image, in which compact radio continuum sources are clearly detected against the background noise. Diffuse Galactic synchrotron emission is largely absent in total intensity because its emission spans angular scales too large to be recovered by the interferometer.
The top-right panel displays the extinction map, tracing the dust distribution and filamentary structure of the molecular cloud. To assess whether the polarized synchrotron features originate within the Perseus molecular cloud or instead arise from foreground Galactic structures, we compare the Stokes $Q$ and $U$ patterns with the dust distribution. If the polarized emission were physically associated with the cloud, one would expect spatial correspondence between polarization structures and regions of high extinction. However, no such alignment is observed. The polarized patterns do not trace the filamentary or high-extinction features seen in the dust map, indicating that the $Q$ and $U$ structures are dominated by foreground magnetized gas unrelated to the Perseus cloud.
The bottom-left and bottom-right panels present the Stokes $Q$ and $U$ mosaics, respectively, both exhibiting broad, wave-like and filamentary patterns characteristic of diffuse Galactic synchrotron emission at L-band frequencies. Similar structures have been reported in other polarization surveys \citep{taylor2003canadian, jelic2014initial, van2017faraday} and are commonly interpreted as tracing coherent magnetic fields in the local ISM on intermediate angular scales. These polarized features are further shaped by Faraday rotation and depolarization effects arising from the magnetized, ionized ISM \citep{wieringa1993small}. Small-scale variations in Faraday rotation along the line-of-sight cause shifts in the polarization angle, producing the alternating bright and dark bands visible in the $Q$ and $U$ maps.

In regions where diffuse Galactic polarized emission is strong, the polarization signal can exhibit Faraday-complex behavior due to fluctuations in the foreground magneto-ionic medium. This causes the polarization angle to vary nonlinearly with wavelength squared, deviating from a simple Faraday-thin model and biasing RM measurements of compact background sources. To account for this effect, we introduce a flag indicating whether a source lies within regions affected by strong Galactic polarized emission; this flag is described in Section~\ref{RM_catalog}.

\subsection{Catalog of Rotation Measures} \label{RM_catalog}

\begin{figure*}[ht!]
    \centering
    \includegraphics[width=1\textwidth]{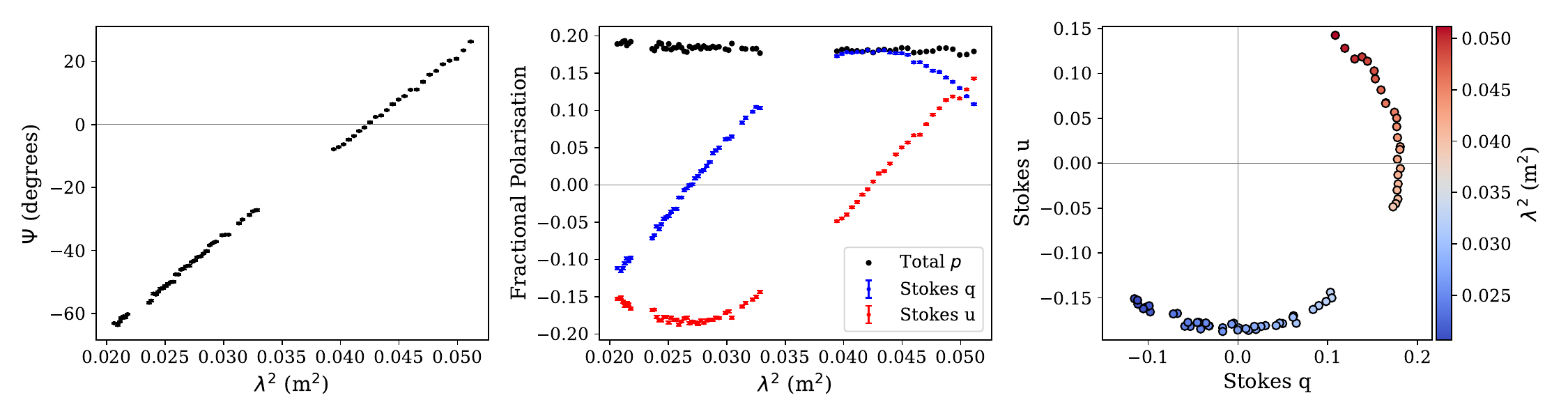}
    \caption{Spectral and polarization properties of a background source used for RM analysis.
    This source exhibits a high fractional polarization with a well-defined and coherent polarization signature.
    \textbf{Left:} The polarization angle ($\Psi$) changes with $\lambda^2$, and the trend is linear, consistent with a single Faraday-thin component.
    \textbf{Middle:} Fractional polarization as a function of $\lambda^2$, with Stokes $q$ and $u$ components fitted.
    \textbf{Right:} Stokes $q$--$u$ diagram showing a smooth circular trajectory, indicative of coherent Faraday rotation.}

    \label{fig:spectra_qu_theta_flux}
\end{figure*}


\begin{figure*}[]
    \centering
    \includegraphics[width=0.71\textwidth, height=0.33\textwidth]{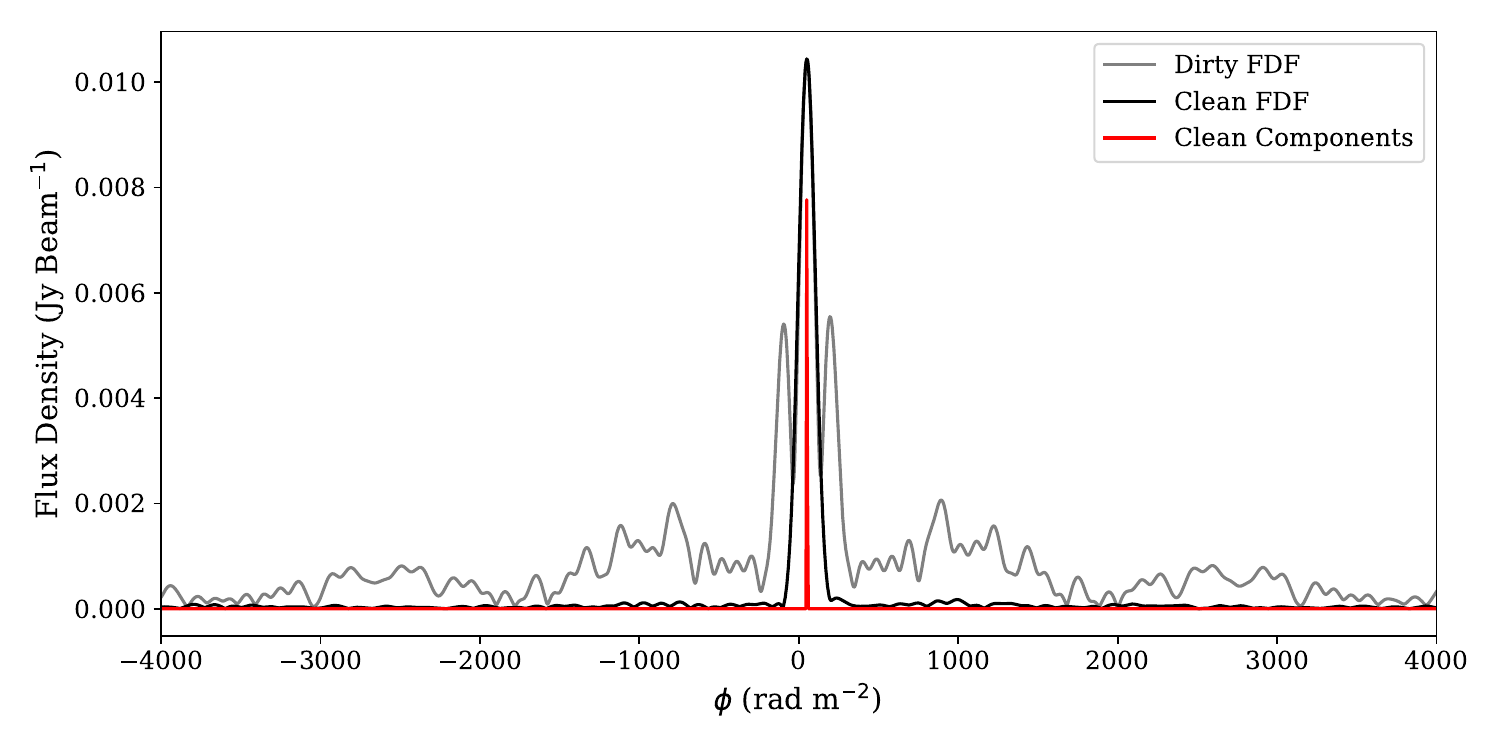}
     \caption{Example of RM CLEAN used on a typical polarized source. The gray line shows the Dirty FDF, while the black line displays the CLEANed FDF after deconvolution. The red curve represents the CLEAN components, highlighting discrete peaks in Faraday depth.}
    \label{fig:rmclean}
\end{figure*}

This section describes the polarization properties of background radio sources in our dataset and outlines the RM Synthesis and RM CLEAN procedures used to obtain accurate RM. We also present the resulting catalog and summarize the statistical behavior of the derived RMs and fractional polarizations.

Figure~\ref{fig:spectra_qu_theta_flux} illustrates the polarization properties of a background source from our dataset governed by the RM Synthesis technique. Each panel shows a key aspect of the source’s behavior across the L-band.
The left panel shows the linear polarization angle $\Psi$ against $\lambda^2$. The clear linear trend is a defining signature of Faraday rotation and represents the presence of a single dominant Faraday-thin component. The slope of this line is the RM.
The middle panel shows the fractional polarization components \(q = Q/I\) (blue) and \(u = U/I\) (red) as a function of \(\lambda^2\). The smooth variation of these components suggests a small amount of depth depolarization across the observed band.
Finally, the right panel displays the Q–U plane trajectory. The helical path traced by the red and blue circles is compatible with a Faraday-rotated signal.


Figure~\ref{fig:rmclean} demonstrates the outcome of the RM CLEAN process. The gray line shows the Dirty FDF, which contains strong side-lobe artifacts due to convolution with the RMSF. The black curve represents the CLEANed (true) FDF after the deconvolution steps, and the red line indicates the CLEAN components recovered during the process. Compared to the original Dirty FDF, the CLEANed version reveals a more compact, isolated peak, characteristic of a Faraday-thin source, and enables a much more accurate measurement of the RM.
Applying RM CLEAN is especially valuable if multiple Faraday components are present along the line-of-sight. 

\begin{table*}
\centering
\small
\renewcommand{\arraystretch}{1.1}
\caption{Example rows from the MRT RM catalog of true Faraday rotation detections toward the Perseus cloud. Columns include source positions, RM measurements, fractional polarization, flux density, local RMS noise, and quality flags (F1, F2). The complete catalog is available online.}
\label{tab:true_detections_final_sample}
\begin{tabular}{lccccccccc}
\toprule
Source ID & RA & DEC & RM & dRM & fracPol & S\_reffreq & RMS & F1 & F2 \\
          & deg & deg & rad m$^{-2}$ & rad m$^{-2}$ & \% & mJy & $\mu$Jy & flag & flag \\
\midrule

VCPMC J031923.8+310143 & 49.84941 & 31.02869 & 33.78 & 0.49 & 5.48 & 0.04 & 110.8 & 0 & 0 \\
VCPMC J031927.6+312115 & 49.86532 & 31.35421 & 58.60 & 1.46 & 7.51 & 0.01 & 36.5 & 0 & 0 \\
VCPMC J032003.0+310733 & 50.01252 & 31.12604 & 75.95 & 0.12 & 5.09 & 0.11 & 147.1 & 0 & 1 \\
VCPMC J032021.4+313824 & 50.08925 & 31.64013 & 47.79 & 0.49 & 6.89 & 0.02 & 47.2 & 0 & 0 \\
VCPMC J032027.7+315638 & 50.11581 & 31.94413 & 44.81 & 1.05 & 15.68 & 0.00 & 30.8 & 0 & 0 \\
VCPMC J032044.6+314601 & 50.18621 & 31.76695 & 54.02 & 0.62 & 6.93 & 0.02 & 37.8 & 0 & 0 \\
VCPMC J032045.4+311206 & 50.18923 & 31.20176 & 46.88 & 0.68 & 9.37 & 0.01 & 47.5 & 0 & 0 \\
VCPMC J032055.6+313602 & 50.23199 & 31.60075 & 49.67 & 1.74 & 19.46 & 0.00 & 30.3 & 0 & 0 \\
VCPMC J032057.1+305615 & 50.23797 & 30.93758 & 54.35 & 2.01 & 8.28 & 0.00 & 32.8 & 0 & 0 \\
VCPMC J032059.3+310139 & 50.24714 & 31.02774 & 45.31 & 2.17 & 2.84 & 0.01 & 34.8 & 1 & 0 \\

\bottomrule
\end{tabular}
\end{table*}


\begin{figure}
    \centering
    \includegraphics[width=0.47\textwidth, height=0.34\textwidth]{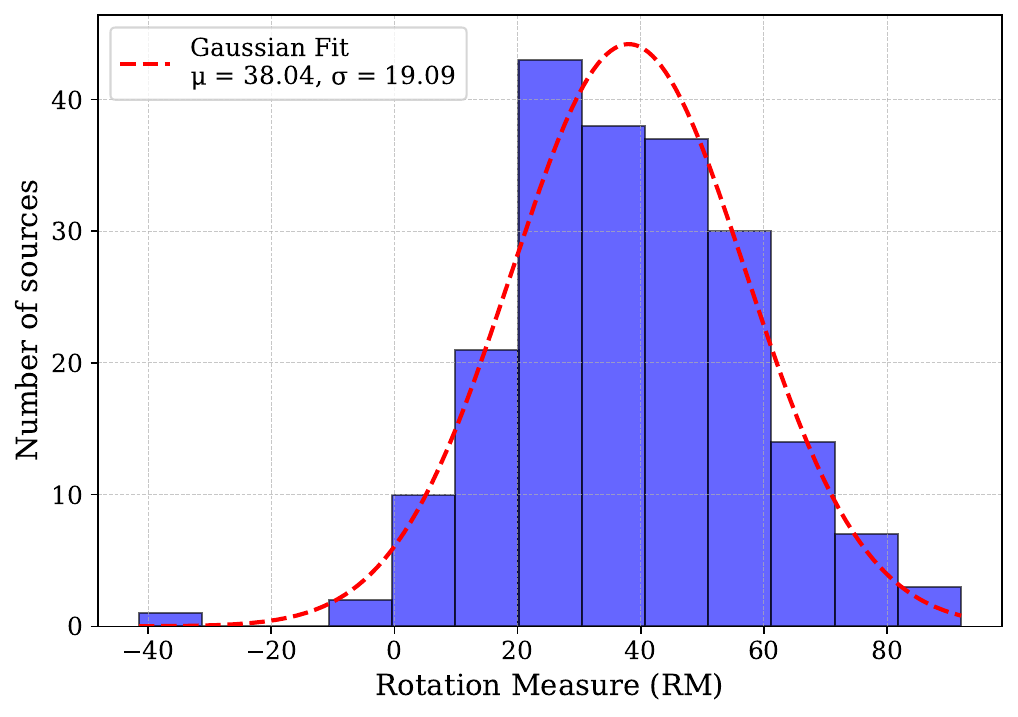}  
    \caption{Histogram of RMs for 205 polarized background sources detected in the Perseus field. The red dashed line shows a Gaussian fit to the data, yielding a mean of \(\mu = 38.04\) rad m\(^{-2}\) and a standard deviation of \(\sigma = 19.09\) rad m\(^{-2}\). This distribution captures the statistical variation in RM across the observed region.}
    \label{fig:RM_hist}
\end{figure}

\begin{figure}
    \centering
    \includegraphics[width=0.5\textwidth, height=0.3\textwidth]{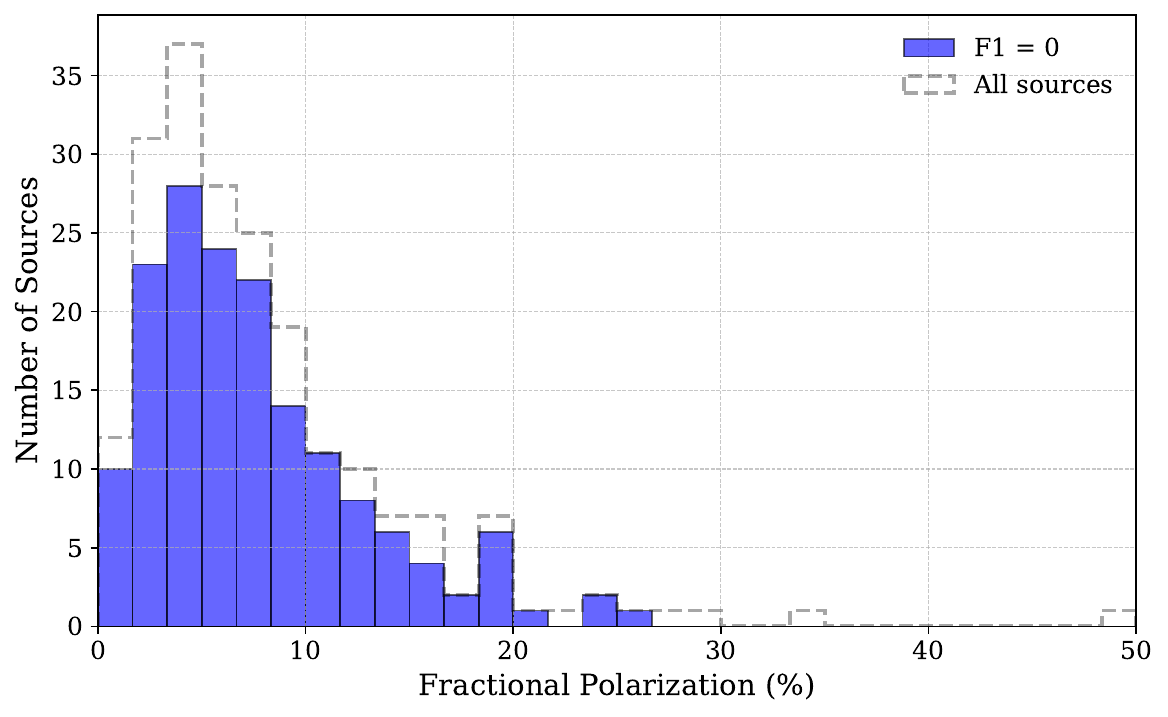}  
    \caption{Histogram of linear fractional polarization for polarized radio sources in the Perseus field. The dashed gray histogram represents all detected polarized sources, and the filled blue histogram shows sources with F1 = 0 (outside regions of diffuse Galactic polarized emission). The exclusion of sources flagged by F1 reduces the high–fractional-polarization tail of the distribution.}
    \label{fig:fracpol_hist}
\end{figure}

Table~\ref{tab:true_detections_final_sample} lists a subset of entries from the MRT catalog of polarized sources. For each source we include the source identifier, equatorial coordinates (RA, DEC), RM with its uncertainty (E\_RM), fractional polarization (fracPol), total flux density at the reference frequency (S\_reffreq), and the local RMS noise measured away from the Faraday peak. 
To identify regions affected by diffuse Galactic polarized emission, we constructed a binary mask from the background linear-polarization intensity ($P$) images for each dataset. Each $P$ map was smoothed with a Gaussian kernel and thresholded at a high percentile of the smoothed distribution to isolate large-scale diffuse features.
The first flag (F1) indicates the location of the source relative to regions of diffuse Galactic polarized emission: 0 for sources outside diffuse emission, 1 for sources inside diffuse emission, and 2 for sources inside diffuse emission where the polarized intensity exceeds twice the maximum value within the diffuse mask. This flag highlights cases where foreground Galactic polarization may bias or dominate the measured RM.
The second flag (F2) characterizes the Faraday complexity of the source based on the structure of the FDF : 0 denotes a single significant Faraday peak, while 1 indicates multiple significant peaks. The F2 classification was determined through visual inspection of the FDF and CLEAN components. Together, these flags allow users of the catalog to distinguish robust RM measurements from sources potentially affected by foreground contamination or Faraday-complex behavior.

The distribution of noise values in our Stokes $Q$ and $U$ maps has a median RMS of $31.2~\mu\mathrm{Jy\,beam^{-1}}$. To ensure reliability, we applied a conservative detection threshold: only sources with RM peak amplitudes exceeding $8\sigma$ the local RMS were retained \citep{george2012detection}. This yields a catalog of 205 high-confidence RM measurements (SNR $> 8$), which underpins our subsequent analysis of the magnetic-field structure in the Perseus region.

Figure~\ref{fig:RM_hist} shows the distribution of RMs derived from the 205 background polarized sources in our 19B dataset. A least-squares fit of a Gaussian yields a mean value of \(\mu = 38.04\) rad m\(^{-2}\) and a standard deviation of \(\sigma = 19.09\) rad m\(^{-2}\). The positive mean RM is consistent with an average magnetic field component directed toward the observer in the Perseus region. This distribution provides a useful statistical characterization of the Faraday rotation environment sampled by our dataset, as it allows us to quantify the typical RM values and their spread across the field.

To estimate the large-scale Galactic contribution along this line-of-sight, we use the all-sky Galactic RM reconstruction of \citet{hutschenreuter2022galactic}. At the position of the Perseus molecular cloud, their model predicts a foreground value of $\mathrm{RM}_{\rm fg} \simeq 48~\mathrm{rad~m^{-2}}$, comparable to the mean RM of our sample, indicating that a substantial fraction of the observed Faraday rotation arises in the Galactic foreground. A detailed treatment of this component, including its subtraction when inferring the RM associated with the Perseus cloud, is currently underway and will be presented in a forthcoming study (Hajizadeh et al., in prep.).

Figure~\ref{fig:fracpol_hist} shows the distribution of linear fractional polarization for polarized sources in our catalog. The dashed gray histogram represents all detected polarized sources, while the filled histogram shows the subset of sources with F1 = 0, corresponding to sources located outside regions affected by diffuse Galactic polarized emission. The exclusion of sources flagged by F1 removes a portion of the high–fractional-polarization tail, indicating that some sources with elevated fractional polarization are associated with regions of strong diffuse Galactic polarized emission rather than intrinsic source polarization.

\begin{figure*}
    \centering
    \includegraphics[width=0.9\textwidth, height=0.6\textwidth]{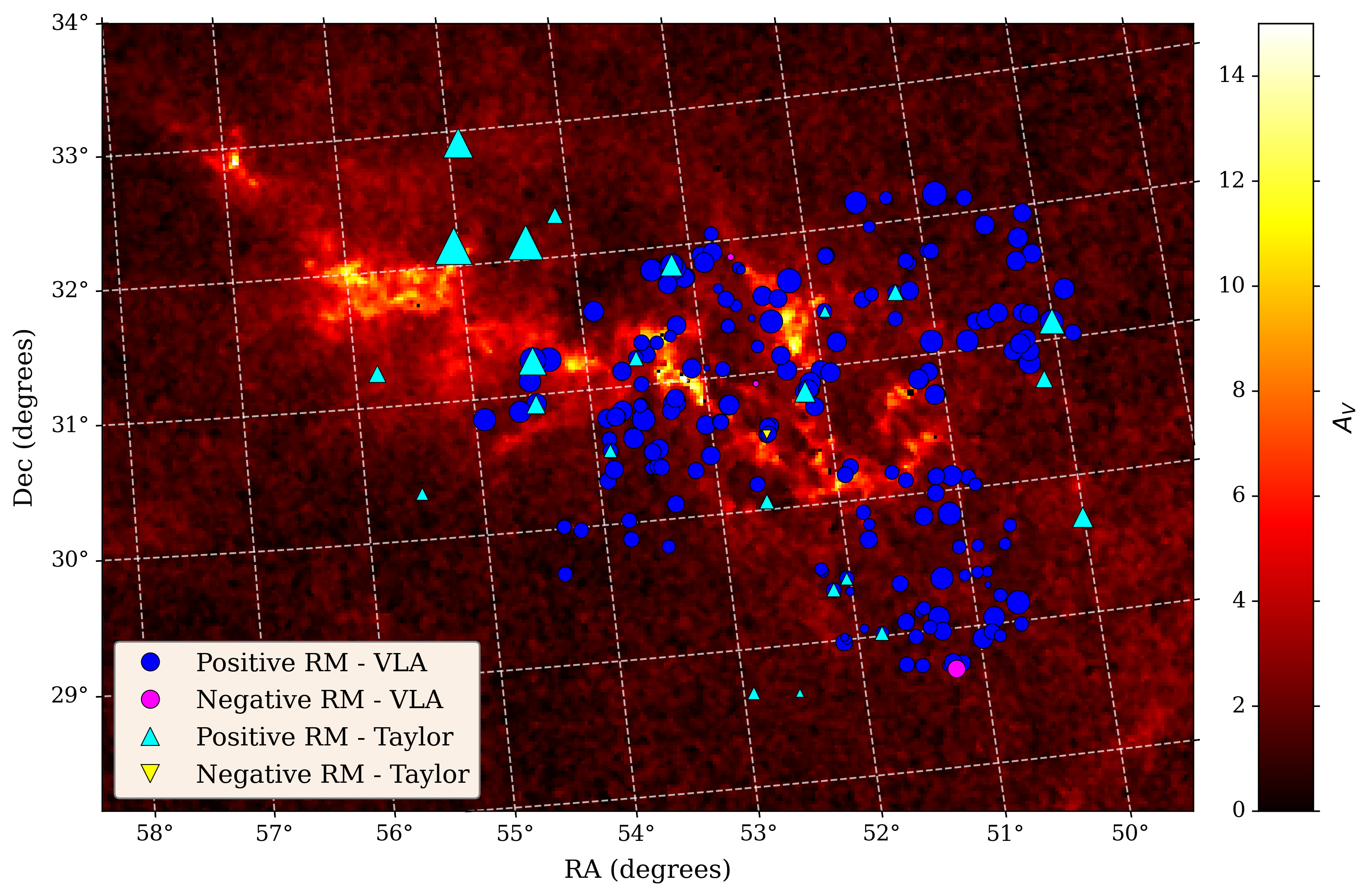}  
    \caption{Extinction map ($A_V$) of the Perseus molecular cloud overlaid with RM detections from this work and from the catalog of \citet{taylor2009rotation}. The color scale represents $A_V$, tracing dust column density, with brighter regions indicating higher extinction. Blue and magenta circles denote sources from this work with positive and negative RM values, respectively, while cyan and yellow triangles indicate positive and negative RM values from the \citet{taylor2009rotation} catalog. The size of each marker is proportional to the absolute value of the RM, illustrating variations in Faraday rotation across the field. The spatial distribution of RMs indicates how complicated the magneto-ionic structure is inside and surrounding the molecular cloud.}

    \label{fig:RM_map}
\end{figure*}

One of the most important results of our work is that the source density per deg$^2$ of RMs across the Perseus cloud has greatly improved because our VLA L-band observations are more sensitive and cover a wider range of frequencies. Figure~\ref{fig:RM_map} highlights this improvement by comparing our new RM detections, shown as blue and magenta circles, with those from the NVSS-based catalog of \citet{taylor2009rotation} represented with cyan and yellow triangles. The increase in the number of RM measurements, along with their improved precision, demonstrates the potential of broadband polarization surveys in probing the molecular clouds. Our new dataset allows for a significantly denser and finer-grained map of $B_\parallel$ than previously available. The six cyan triangles on the far left of the figure correspond to sources outside the current dataset; this region will be included in the upcoming 24A observations in our future work.

Across the 13.8~deg$^2$ area covered by our VLA observations, we identified 205 reliable RM detections, corresponding to an RM density of approximately 14.81~sources per deg$^{2}$. This represents more than a tenfold increase compared to the NVSS RM catalog, which provides a typical density of about 1.25~sources~deg$^{-2}$ in the same region \citep{taylor2009rotation}.


\section{Conclusions and Future Work} \label{sec:conclusion}

We have presented a comprehensive, wideband polarimetric study of extragalactic radio sources behind the Perseus molecular cloud based on VLA L-band observations. This work produced both a detailed radio source catalog and a high-resolution Faraday RM map of the region. From our mosaicked field, we identified 1410 radio sources, for which we provide positions, integrated flux densities, and spectral indices computed across nine SPWs. Most sources exhibit negative spectral indices, consistent with synchrotron emission, supporting their use as background probes of the magnetized ISM.

Among these, we detected 205 polarized sources with significant RM measurements, extracted using RM Synthesis and RM CLEAN. This RM sample increases the density of Faraday rotation measurements to approximately 14.81~deg$^{-2}$ across $\sim13.8$~deg$^2$, representing a more than tenfold improvement compared to the NVSS-based catalog of \citet{taylor2009rotation}. Our RM detections exhibit a mean of $38.04~\mathrm{rad~m^{-2}}$ with a standard deviation of $19.09~\mathrm{rad~m^{-2}}$, and display smooth polarization angle rotation as a function of $\lambda^2$, indicative of Faraday-thin behavior. 
The high density of RM measurements achieved in this study provides a powerful new probe of the line-of-sight magnetic field component, $B_\parallel$, across the Perseus molecular cloud. This enhanced sampling enables the identification of coherent large-scale variations in rotation measure across the cloud, with additional localized structure superimposed on smaller angular scales.

We checked our source catalog, spectral index measurements, and RM values against NVSS to make sure that the flux scales were the same and that the RM measurements were better than those that used data with limited bandwidth and lower sensitivity. The resulting catalogs provide a robust foundation for future studies of magnetic field morphology and its role in regulating gas dynamics and star formation in the ISM.

In future work, we will use these high-resolution RM measurements to obtain $B_\parallel$ and, when we add them to existing dust polarization data, we get a better picture of the cloud's three-dimensional magnetic field shape \citep{tahani20223d}. These findings give us important information about how magnetic fields affect the growth of molecular clouds and the processes that control star formation.

Our 2024 follow-up observations will expand the mosaic coverage and are expected to significantly increase the polarized source count, further refining RM sampling. Together with the present work, they will provide one of the most detailed Faraday RM maps of any nearby molecular cloud to date, establishing Perseus as a benchmark region for future studies of magnetized star-forming environments.

\section{Software}

\software{
CASA \citep{mcmullin2007casa, casa2022casa},
PyBDSF \citep{mohan2015pybdsf},
RM-Tools/RM-CLEAN \citep{purcell2020rm},
Astropy \citep{astropy2022astropy}, 
lmfit \citep{newville2016lmfit}
}

\section{Acknowledgment}
This work makes use of observations obtained with the Karl G. Jansky Very Large Array, 
operated by the National Radio Astronomy Observatory (NRAO), a facility of the National Science Foundation 
operated under a cooperative agreement by Associated Universities, Inc. 
HH acknowledges the use of AI-assisted tools for language refinement and code verification during manuscript preparation.
M.T. acknowledges Emanuel Momjian for providing helpful input and the NRAO help desk for their assistance with the three VLA proposals submitted (19B-053, 24A-376) for this project; Wednesday Lupypciw for contributing to initial data examination; Cameron Van Eck for helping select pyBDSF from various source detection algorithms and assisting with RM Synthesis questions; Trey Wenger for discussing the data and suggesting the use of self-calibration when other approaches did not work well; and particularly CANFAR/CADC for providing storage space for this project. M.T. was supported by Banting, Covington (National Research Council Canada), and KIPAC fellowships during this project. 

\bibliographystyle{aasjournal}
\bibliography{main.bib}  

\end{document}